%% file: main.tex
\documentclass{article}

\usepackage[preprint]{neurips_2026}

\usepackage[utf8]{inputenc}
\usepackage[T1]{fontenc}
\usepackage[hidelinks]{hyperref}
\usepackage{url}
\usepackage{amsmath}
\usepackage{microtype}
\usepackage{xcolor}
\usepackage{graphicx}

\input{paper_numbers}

\title{Dutch Books for Language Models}

\author{%
  Isaiah Andrews\\
  MIT and NBER\\
  \texttt{iandrews@mit.edu}
  \And
  Suproteem Sarkar\\
  University of Chicago\\
  \texttt{suproteem@uchicago.edu}
}

\begin{document}

\maketitle

\begin{abstract}
People increasingly use language models to support life decisions. Many such decisions involve a probabilistic forecast: How likely is a major life event, a natural disaster, or an economic outcome? Users of language models may implicitly trust that these forecasts fall out of a coherent world model. In this paper, we evaluate the coherence of language model probabilistic forecasts through a procedure that builds on a theorem due to de Finetti. We elicit forecasts from language models across events generated from stock returns data. We then use linear programs to compute the largest Dutch-book profit - the profit an arbitrageur could guarantee by betting against model-generated probabilities - which we use as a measure of incoherence. Our procedure does not require outcome labels, so we can evaluate coherence even in settings where outcomes are not observed or have not yet resolved. We find substantial evidence of incoherence in language model forecasts. Such incoherence increases when there are richer logical relationships between events, and irrelevant contextual details can increase incoherence by an order of magnitude. We conclude by discussing how alternative training strategies may improve probabilistic coherence.
\end{abstract}

\section{Introduction}\label{sec:intro}

Language models are increasingly used for probabilistic forecasting, and the performance of model ensembles rivals that of groups of humans \citep{schoenegger2024, halawi2024, alur2025}.  At the same time, model predictions are sometimes incoherent, for instance failing to respect that the probability of an event and its complement must sum to one \citep{paleka2024}.

A classic theorem due to de Finetti [\citeyear{definetti1937}, \citeyear{definetti1974}] establishes that probabilistic forecasts for a finite collection of events are coherent (that is, consistent with some probability distribution) if and only if an arbitrageur facing odds based on those probabilities has no Dutch book, that is, no way of placing bets which yields a profit regardless of the outcome.  This provides an immediate route to check the coherence of a group of forecasts: one may simply play the role of an arbitrageur aiming to maximize the worst-case profits, treating stated probabilities as prices. Zero profits indicate coherence. Importantly, this check requires no outcome labels, since it is based on the probabilistic forecasts (and their known logical structure) alone.  \citet{andrews2026} proposes this use of de Finetti's result in language model evaluation and training, while this paper explores the question empirically, importing no-arbitrage pricing as a label-free coherence measure.

Our empirical evaluation of probabilistic coherence in language models uses stock market data.  In particular, we ask models to predict the probability that future stock returns fall in a given return bin, based on recent return history and in some cases additional information.  Considering multiple such bins, as well as their unions and complements, across days and across assets generates a collection of events whose logical structure is known by construction.  We then score coherence for predictions on (a subset of) these events by the largest profit that an arbitrageur could guarantee by betting against the models' stated probabilities.  This choice of domain helps tie our prediction exercise to practice, since binary option contracts (i.e., contracts which pay off a fixed amount if some event occurs, and nothing otherwise), including contracts based on stock returns, are common in prediction markets.  Moreover, prediction market probabilities are routinely interpreted as probabilistic forecasts \citep{wolfers2004, wolfers2006, snowberg2013, diercks2026}. The model predictions we study are not tradable, however, so the ``profits'' we consider are measures of the degree of incoherence rather than financial returns.

Our evaluation covers 100 stock-days, 15 models, and a variety of elicitation arms detailed below, with {365,100} total elicitations. Our main contributions are as follows:
\begin{itemize}
    \item We develop an \textbf{evaluation of probabilistic coherence in language model forecasts}. Our method is exhaustive with respect to a fixed event set, and does not require outcome labels.
    \item We find that \textbf{incoherence, as measured through arbitrage profit, is common and often sizable}. We find larger variation in coherence than in accuracy.
    \item We find that \textbf{forecasts that involve rich logical relationships have higher incoherence}, even for models that are more coherent on simpler forecasting tasks.
    \item We show that coherence varies strongly with the elicitation protocol. \textbf{Irrelevant contextual details can increase incoherence} by an order of magnitude.
\end{itemize}
In addition, we discuss training strategies that may improve coherence in language model forecasts. 

The closest work to ours is \citet{paleka2024}, who propose to check selected cross-forecast consistency properties for model forecasts, and show that incoherent forecasts can be improved (on average across events) regardless of the outcome.  We offer a similarly arbitrage-based coherence measure, but differ in that we focus on exhausting the implications of probabilistic coherence (and thus use a global coherence measure) for a given set of forecasts. Like \citet{paleka2024}, however, we find that across models coherence is positively correlated with accuracy.  Incoherent probability judgments are well documented in humans and, more recently, in models: \citet{betz2023} and \citet{fluri2024evaluating} measure probabilistic consistency of language models and explore methods to improve it, while other work finds incoherent probability judgments \citep{zhu2024} and proposes repairs for Dutch-book and related vulnerabilities \citep{chadwick2025, matta2026rethinking}.  Our elicitation approach relates to work on eliciting model confidence \citep{kadavath2022, lin2022, tian2023}, improving logical agreement among assertions \citep{kassner2021, mitchell2022}, as well as work measuring sensitivity to example order \citep{lu2022}, prompt format \citep{sclar2024}, and query protocol \citep{lei2026}. More broadly, our results relate to a growing literature on world models \citep{li2023emergent,vafa2024world,vafa2025foundation}. Unlike calibration \citep{guo2017}, proper scoring rules \citep{gneiting2007}, and recent benchmarks that score model probabilities on prediction-market or platform questions \citep{nel2025, cheng2026, karger2025}, coherence requires no realized outcomes.

\section{Coherence and Dutch Books}\label{sec:setup}

Consider a finite collection of events $E_1,\ldots,E_n$ whose logical relationships are known.  The algebra these events generate (the collection of all events expressible through complements, intersections, and unions) partitions the possible outcomes into atoms $\omega_1,\ldots,\omega_m$. These atoms are the finest outcomes that the events jointly distinguish.  Let the incidence matrix $M\in\{0,1\}^{n\times m}$ record the mapping of atoms to events, where $M_{ij}=1$ if atom $\omega_j$ belongs to event $E_i$, and let $p\in[0,1]^n$ denote the probabilities a model states for the events.

\begin{figure}[!t]
  \centering
  \includegraphics[width=\linewidth]{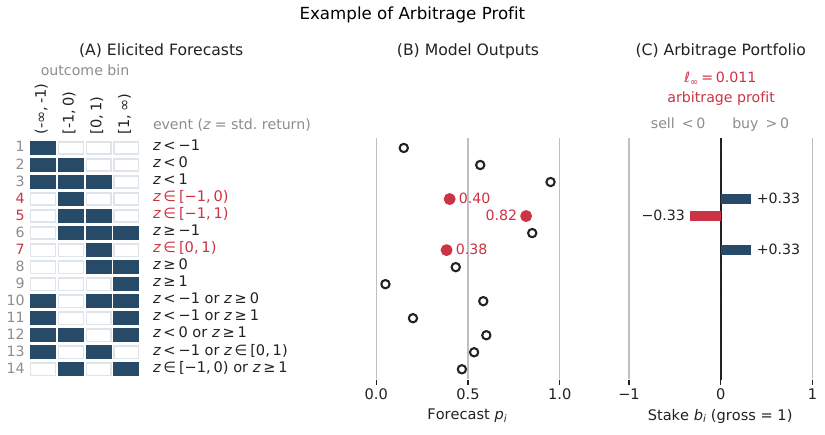}
  \caption{\textbf{An example of arbitrage profit.} The figure shows, for a particularly incoherent draw, the queried events (left), the model's stated probabilities (center), and the corresponding portfolio (right). The three binding events are highlighted. Any coherent forecast satisfies $P(-1 \leq z < 0)+P(0 \leq z < 1)= P(-1 \leq z < 1)$. In this case the model's stated probabilities are $0.4000$ and $0.3833$ and $0.8167$, respectively. Therefore, buying one third unit each of events 4 and 7 and selling one third unit of event 5 yields profit $0.01113$ on every atom. In the figure, numbers are rounded to two significant digits.}
  \label{fig:certificate}
\end{figure}

Consider an arbitrageur placing bets at the model's stated probabilities.  A bet of stake $b_i$ on event $E_i$ costs $b_{i}p_{i}$ and pays $b_i$ if the event occurs, so its net payoff on atom $\omega_j$ is  $b_{i}(M_{ij}-p_{i})$. Negative stakes correspond to bets against the event.  A \emph{Dutch book} against $p$ is a collection of bets whose net payoff is strictly positive no matter which atom occurs, and our measure of incoherence is the profit an arbitrageur can guarantee per unit of total stake,
\begin{equation}\label{eq:linf}
    \ell_{\infty}(p)=\max_{\left\Vert b\right\Vert _{1}\leq1}\;\min_{j=1,\ldots,m}\;\sum_{i=1}^{n}b_{i}\left(M_{ij}-p_{i}\right).
\end{equation}
The constraint $\left\Vert b\right\Vert _{1}=\sum_{i}|b_{i}|\leq1$ caps the (notional) amount wagered, and ensures that the profit is finite.  Standard duality arguments, reviewed in Appendix \ref{app:duality}, show that our incoherence measure equals $\min_{\pi\in\Delta^{m-1}}\left\Vert M\pi-p\right\Vert _{\infty}$, the sup-norm distance from $p$ to the set of coherent forecasts, for $\Delta^{m-1}$ the set of probability distributions over the atoms.  We accordingly call $\ell_{\infty}(p)$ the \emph{arbitrage profit}.  From the dual form it is immediate that $\ell_{\infty}(p)=0$ exactly when the stated probabilities are coherent, meaning that $p$ extends to some probability measure over the algebra, recovering the characterization of de Finetti [\citeyear{definetti1937}, \citeyear{definetti1974}]. Figure \ref{fig:certificate} works through an example.

We emphasize three properties of our coherence measure.  First, coherence is a property of model responses, which may or may not correspond to ``true'' model beliefs in a meaningful sense.  Second, coherence is not accuracy: a forecaster who assigns probability one to a single well-defined outcome (e.g. all stock prices rise by precisely 150\%) is coherent no matter what events we consider, so while dogmatism may make for poor forecast accuracy, it is perfectly coherent.  Third, our coherence measure depends on the events queried, and $\ell_{\infty}$ is weakly increasing as events are added to a set.

\section{Data and Experimental Design}\label{sec:design}

\paragraph{Sample.} We build a stock-date sample of past, current, and future returns matched to current news headlines. The returns data come from CRSP and headline data from Refinitiv. We draw the sample as follows. First, we consider stocks $s$ and anchor dates $t$ in the CRSP daily stock file with $t$ ranging from August 15, 2025 to December 23, 2025. We choose this window as it falls after the August 5, 2025 release date of GPT-OSS-120B, which is the workhorse model for much of our analysis, allowing us to limit lookahead bias from pretraining \citep{glasserman2023, sarkar2024}. We keep stock-date observations with at least 60 trading days of prior return history, at least five subsequent trading days of return history, and for which a matched news headline in the Refinitiv data was published before the NYSE trading day close on date $t$.  We then draw 10,000 stock-days uniformly without replacement.  From this 10,000-observation sample, we sample 50 trading dates uniformly at random and on each date uniformly draw two stocks. This sampling strategy yields 98 distinct stocks, two of which appear on two dates each. These design choices give us a consistent sample to use across our extension experiments, including providing models with news headlines and evaluating model outputs for two stocks on the same day.

\paragraph{Models.} We evaluate fifteen models, drawn from a range of open- and closed-weight families - Appendix \ref{app:accounting} lists providers and additional details.  Prompts are identical across models,  we use each provider's default reasoning settings, set temperature equal to one if the provider accepts the parameter (again, see Appendix \ref{app:accounting} for details), and set a 8,192 token completion limit. Early exploration on other data suggested that GPT-OSS-120B was particularly coherent, so we have taken it as the workhorse of our analysis, e.g., when exploring the extent to which coherence varies with the details of the elicitation or the richness of the events queried. This model is particularly attractive as our workhorse as it was released before our sample period, which reduces the potential for lookahead bias, and has open weights, which allows for greater replicability compared to closed-weight models \cite[cf.][]{chen2024chatgpt}.

\paragraph{Event panels.} For horizons $h\in\{1,\ldots,4\},$ we consider the events that the return from day $t+d-1$ to day $t+d$ for $d \in \{1, \ldots, h\}$, divided by the standard deviation of returns over the 60 trading days preceding the anchor date $t$, falls in one of the four bins with edges at $-1,0$ and $1.$ This coarsening yields $4^h$ possible outcomes at horizon $h$. At horizon one we consider all 14 events expressible based on the four bins (excluding the empty and certain events).  For longer horizons, we add the 14 non-trivial events for each day, as well as additional events spanning multiple days. Specifically, for each pair of days  we coarsen the bins into up versus down, tails versus center, and each bin versus the rest, and within each coarsening elicit the probability of the four intersections across the two days. Finally, for $h\ge3,$ we include the up-versus-down patterns for each triple of days, and at $h=4$, also across all four days. By construction this yields nested event panels: every event in the horizon $h$ panel appears in the horizon $h+1$ panel as well.

\paragraph{Elicitation.} In all elicitations, we provide the model with each stock's previous 60 daily returns, normalized to have standard deviation one, as well as the standard deviation over those 60 days, with real dates shown and the stock's identity withheld except where otherwise noted. Our baseline elicitation protocol asks about a single event in each query, with each query submitted to an independent session and no access to other queries or their answers, and a fixed instruction identifying the model as a forecaster.  We use this one-query-at-a-time format to mimic deployments where different questions arrive from users in sessions with no shared context.

Our baseline elicitations ask about the 14 horizon one events for a single stock.  We also consider elicitations that ask about two stocks at once, and others that consider longer horizons as described above. We further consider several variants of our baseline elicitation, each changing one aspect at a time: arms which add the stock's bin frequencies over the 60 displayed days, or additional guidance on a forecasting procedure to use; an identity arm which names the company and a headline arm further adding the stock-day's matched headline; grouped arms which ask about multiple events at once, ranging from logically unrelated events to the full event panel; a red herring arm which varies return-irrelevant aspects of the prompt, and averaging arms which average across multiple elicited responses before assessing coherence.

For all analyses, except where noted otherwise, we run five passes for each stock-day prompt, compute the LP solution for each pass separately, and use the average LP value across the five passes for each stock-day prompt. We refer to this statistic as the arbitrage profit and use it as a measure of incoherence. Appendix \ref{app:materials} provides additional details.

\paragraph{Inference.} Except where noted otherwise, all confidence intervals are 95 percent percentile bootstrap intervals, clustered on date, based on 20,000 replicates.  Clustering on date captures the randomness arising from our random sampling of dates and, conservatively, the sampling of stocks within each date.

\section{Results}\label{sec:results}

\subsection{Language Model Forecasts Are Incoherent}\label{sec:prevalence}

We find evidence of incoherence in language model forecasts. We first report results from the baseline elicitation. For each of the 100 stock-days, we elicited forecasts from GPT-OSS-120B for each of the 14 questions over horizon $h=1$. The mean profit was \numBaselineHOneFivePassProfitMean{} per unit of gross stake (95\% CI [\numBaselineHOneFivePassProfitCiLow{}, \numBaselineHOneFivePassProfitCiHigh{}]). We observed a profit of at least 0.1 percent per unit stake in \numNBaselineHOneStockDaysProfitAboveTenBps{} of 100 stock-days, at least 0.5 percent in \numNBaselineHOneStockDaysProfitAboveFiftyBps{}, and at least 1 percent in \numNBaselineHOneStockDaysProfitAboveOneHundredBps{}. Such incoherence exceeds what could be mechanically generated by rounding, since we instruct models to report outputs to four decimal places, which can produce at most $5\times10^{-5}$ of arbitrage profit per unit of gross stake.

\subsection{Incoherence Varies More Than Accuracy}\label{sec:models}

We find that incoherence exists across families of language models. We extend our analysis by running the baseline elicitation of each of the 14 one-day forecasts across 14 additional models.\footnote{As the knowledge cutoffs for several of these additional models fall after our sample window, lookahead bias from pretraining may influence the accuracy scores for models other than our workhorse model GPT-OSS-120B. Nevertheless, our incoherence metric remains well defined, and would be zero if models had memorized the outcome.} To manage API costs, we run one pass across each of the non-workhorse models instead of taking the average value across five passes. This statistic still targets the same estimand as our workhorse elicitation. For each model, we compute mean arbitrage profit and mean Brier score (the mean squared difference of the indicator for whether an event occurred and its stated probability) across stock-days. Arbitrage profit varies by approximately a factor of 100, from \numBaselineHOneFivePassProfitMean{} for GPT-OSS-120B to \numCrossModelMistralSmallThreePointTwoProfitMean{} for Mistral Small 3.2. Brier scores occupy a much narrower range, from \numCrossModelClaudeOpusFiveBrierMean{} to \numCrossModelMistralSmallThreePointTwoBrierMean{}. The two rankings nevertheless broadly agree: their Spearman rank correlation is \numCrossModelProfitBrierSpearmanCorrelation{}. Thus models with more coherent forecasts also tend to be more accurate, though coherence differences are stronger, even relative to sampling uncertainty: of the 105 possible pairwise model comparisons, \numNModelPairProfitCisExcludeZeroUnadjusted{} have 95 percent confidence intervals for profit differences that exclude zero, compared with \numNModelPairBrierCisExcludeZeroUnadjusted{} for Brier score differences.

\begin{figure}[tb]
  \centering
  \includegraphics[width=\linewidth]{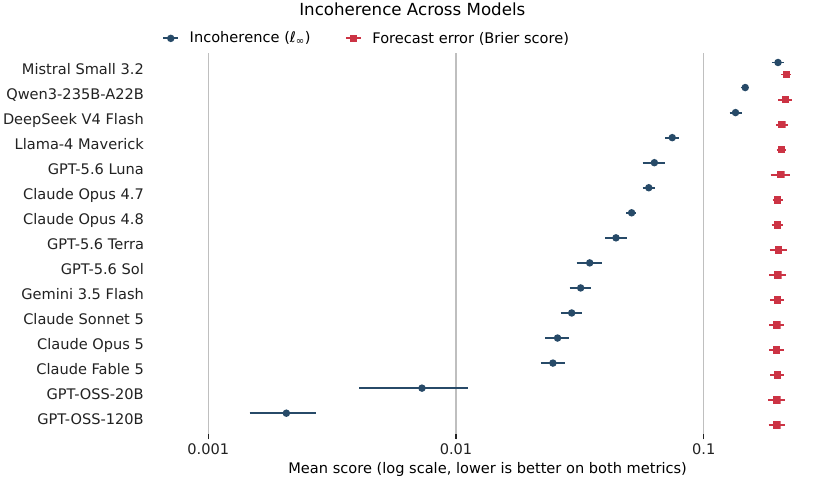}
  \caption{\textbf{Incoherence varies across the 15 evaluated models.} This figure reports mean profit per unit stake and mean Brier score for each model. Error bars corresponding to 95 percent date-clustered bootstrap intervals are reported for each statistic.}
  \label{fig:models}
\end{figure}

\subsection{Questions With Logical Dependencies Generate Additional Incoherence}\label{sec:scope}

\begin{figure}[tb]
  \centering
  \includegraphics[width=\linewidth]{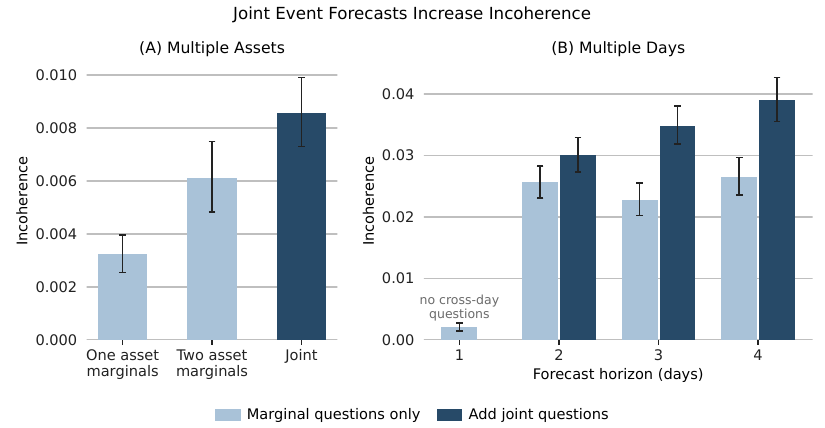}
  \caption{\textbf{Questions with logical dependencies increase incoherence.} This figure reports mean profit per unit stake across experiments that vary the number of days and assets evaluated per query, as well as whether the queries link multiple days or assets. Panel (A) reports results across multiple assets, and Panel (B) reports results across forecast horizons.}
  \label{fig:horizons}
\end{figure}

\paragraph{Multiple assets.} Questions linking two stocks raise incoherence. For each of the 50 dates in the sample, we provide the previous 60 days of returns for both stocks (``stock A'' and ``stock B'') and then elicit the probabilities assigned to the fourteen one-day-ahead events for each stock separately, as well as the sixteen joint-cell events for the two stocks jointly.\footnote{For symmetry, we run each elicitation twice, swapping the roles A and B, and average the resulting arbitrage profits.}  We then compute the mean arbitrage profits from using each stock's marginal events only, the two sets of marginal events only, and then the full set of events - see panel (A) in Figure \ref{fig:horizons} for the results. When we use the marginal events for one stock at a time, arbitrage profits are significantly higher than in the matched baseline elicitation, suggesting that including information about two stocks in the prompt itself raises incoherence even when only one stock's marginal events are considered. Considering the two sets of marginal events yields incoherence equal to the maximum of the two separately, and so increases the average.  Further including the joint events significantly raises incoherence relative to the two marginals (average increase \numTwoAssetFullMinusBothMarginalsProfitMean{}, with confidence interval [\numTwoAssetFullMinusBothMarginalsProfitCiLow{}, \numTwoAssetFullMinusBothMarginalsProfitCiHigh{}]).

\paragraph{Multiple days.} Questions linking multiple days also raise incoherence. Panel (B) of Figure \ref{fig:horizons} shows the average arbitrage profit at different forecast horizons when we consider only the marginal questions for each day (e.g. at horizon two, the marginal questions about day one and day two), versus adding the joint questions connecting across days.  A complication in interpreting these results is that the prompts for forecast horizon $h$ all bear a preamble asking the model to think about returns out to this horizon, which appears to influence performance even when the event queried is closer.\footnote{For instance, the horizon one forecasts are significantly more coherent when the preamble asks the model to consider the next three days, rather than the next two.}  Consequently, these results are more cleanly interpreted by comparing incoherence within horizon.

\paragraph{Matched queries elicitation.} We have shown that incoherence increases when we add joint events linking outcomes across assets or across days.  This comparison, however, varies both the number of events queried and the richness of their logical relationship.  To isolate the second channel, in Appendix \ref{app:dependence} we consider a complementary exercise where we consider events with more or less rich dependence, holding fixed the number of events queried.  We find that richer dependence is associated with lower coherence.

\subsection{Incoherence Varies Across Elicitation Strategies}\label{sec:helping}

\begin{figure}[tb]
  \centering
  \includegraphics[width=\linewidth]{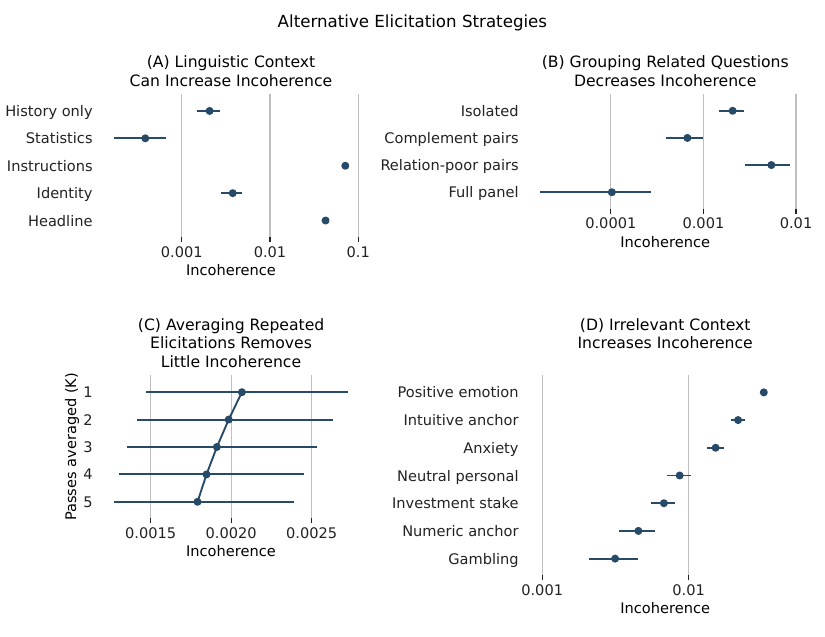}
  \caption{\textbf{Elicitation experiments.} This figure reports mean profit per unit stake across experiments that vary the elicitation procedure. Panel (A) includes additional linguistic context with information about the stock's past returns, identity, or current news, as well as forecasting instructions. Panel (B) elicits multiple probability estimates per query. Panel (C) averages up to five repeated elicitations. Panel (D) adds irrelevant context that should not affect the probability forecast.}
  \label{fig:elicitation}
\end{figure}

Coherence is sensitive to the details of the prompt and elicitation strategy, even holding the set of events fixed.  Figure \ref{fig:elicitation} collects the results from a variety of elicitation experiments, where in each case we consider next-day returns for a single asset.

\paragraph{Information.} Our first experiment varies the information included in the prompt.  In the ``statistics'' arm, we append the bin frequencies over the 60 days already displayed.  In the ``instructions'' arm we specify a forecasting procedure (specifically, count how many of the displayed returns fall in each bin, use the resulting proportions as starting estimates for those bins on each future day, and then apply judgment).  In the ``identity'' arm we disclose the identity of the company, and in the ``headline'' arm we further reveal a news headline about that stock, from before the market close that day.

The results are shown in Panel (A) of Figure \ref{fig:elicitation}. We find that the ``statistics'' arm leads to a statistically significant decrease in arbitrage profits, while instructions lead to a large and highly significant increase.  The identity arm leads to a smaller but statistically significant increase, while the headline arm again leads to a large and significant increase.  Together, the statistics and instructions arms highlight that coherence is sensitive to some changes to the prompt.  Similarly, the identity and headline arms show that adding more information to the prompt may hurt coherence.

\paragraph{Grouping.} Our second experiment varies how many, and which, events are asked about in a given query.  In the ``complement pairs'' arm, each query asks both about an event and its complement.  By contrast, the ``relation-poor pairs'' arm groups pairs of events which are non-nested, with five overlapping pairs and two pairs formed from disjoint events.  Finally, the ``full panel'' arm asks about all 14 next-day events in a single query.

Panel (B) of Figure \ref{fig:elicitation} reports the results. Complement pairing leads to a statistically significant decrease in mean arbitrage profits, while relation-poor pairing leads to a statistically significant increase.  Interestingly, asking about the full panel at once drives incoherence nearly to zero, with the remaining incoherence on the same order as one would expect due to rounding.  Thus, we see that for GPT-OSS-120B and this panel of questions, providing the full event panel suffices to eliminate most of the incoherence. This result relates to the findings of \citet{zhang2026verbalized}, who study how models can be elicited to generate multiple outcomes. We do not expect general users to query language models for forecasts using such an elicitation strategy, and we find strong evidence of incoherence with single-event queries. However, this result does suggest that alternative post-training strategies which encourage models to reason through related events \citep[e.g.][]{puri2026} might increase coherence for single-event queries.

\paragraph{Repeated elicitation.} Our third experiment explores the effect of averaging over multiple responses.  Since $\ell_{\infty}(p)$ is convex in $p,$ we know that computing the arbitrage profit based on the average $p$ over multiple elicitations yields weakly lower profit than computing arbitrage profits separately for each elicitation and then averaging. Moreover, given the randomness inherent in model responses, it seems plausible that the inequality may be strict.  To explore the impact of averaging, we compute arbitrage profits after first averaging responses over 1, 2, 3, 4, and 5 elicitations.

Panel (C) of Figure \ref{fig:elicitation} shows the results.  As expected, averaging improves coherence, though the gains are modest.  The reason appears to be that there is simply not much randomness to remove, since 95 percent of answers are identical across all five passes.  Hence, for GPT-OSS-120B, the incoherence we find does not appear to be primarily driven by the randomness in model responses.

\paragraph{Irrelevant context.} Our final experiment considers seven arms which insert into the query a short note from the requester which arguably should not affect the forecast, ranging from the fact that the requester had coffee that morning, through notes conveying positive or negative emotion about the stock, to numeric anchors.

Panel (D) of Figure \ref{fig:elicitation} reports mean arbitrage profit for each irrelevant-context arm. Statistical significance is assessed using paired differences relative to the baseline elicitation.  We find that most of the arms lead to statistically significant increases in incoherence, with the largest increases generated by the expression of a direct, intuitive probability forecast (``{I don't know much about forecasting, but} my gut says the probability is around 70\%'') and positive emotion (``I'm feeling really optimistic about this''). Other significant arms include ones expressing anxiety (``I'm honestly pretty scared about what this stock might do next''), unrelated numeric anchors (``My phone battery is at 70\%, by the way''), unrelated personal information (``I had coffee this morning, by the way''), and personal investment exposure (``I've got about 10\% of my savings invested in this company, by the way''). While some of these cases arguably convey information about the stock (e.g., positive emotion about a stock might be based on something) most of them should still lead to coherent beliefs across events.

\section{Discussion and Limitations}\label{sec:discussion}

Applying a comprehensive, arbitrage-based measure of forecast coherence for a given set of events, we have shown that probabilistic coherence varies considerably across recently released language models.  Using the open-weight model GPT-OSS-120B, we have further shown that incoherence is larger in more demanding forecasting settings (e.g., more days, more assets) and is responsive to details of the prompt, for instance falling in some cases where additional information is provided.

These results are based on a particular forecasting environment (predicting variance-normalized stock returns) and so do not provide direct evidence on the generalizability of our findings to other domains.  Moreover, our elicitation experiments are based on GPT-OSS-120B, and we have not shown that the coherence ordering we find across models is invariant to the elicitation conditions.

One advantage of our metric is that it does not require outcome labels. This means that coherence can be evaluated even in settings where accuracy cannot be, for example for events involving unknown outcomes. In Appendix~\ref{app:materials}, we include examples applying coherence evaluations to events that have not yet resolved.

We think that these coherence issues are addressable, but may require different post-training strategies. Our results on joint elicitation suggest that models may be capable of making more coherent forecasts when induced to reason across events. This finding suggests that training strategies that encourage test time computation across events may improve coherence. Process reward models are one step toward encouraging such computation. So are RL objectives over groups \citep{orney2026polyeppo,pres2026selfconsistency}, in lieu of example-level training rewards that do not necessarily encourage computation over related states. The Dutch Book linear program may be attractive for such training as it provides an exact set-level reward, which may reduce variance compared to a learned reward model.

\bibliographystyle{plainnat}
\bibliography{references}

\appendix

\section{Duality}\label{app:duality}

This appendix derives the equivalence, stated in Section \ref{sec:setup}, between the arbitrage profit and the sup-norm distance to the set of coherent forecasts.  

Under the setting described in Section \ref{sec:setup}, and for $\pi\in\Delta^{m-1}$ a distribution over the atoms, the expected payoff for a vector of bets $b=(b_1,\ldots,b_n)'$ under $\pi$ is $b'(M\pi-p)$.  Since this is linear in $\pi,$ the minimum over $\pi$ is attained at a vertex of $\Delta^{m-1}$, so the arbitrage profit is equal to
\[
\max_{\left\Vert b\right\Vert _{1}\leq1}\;\min_{\pi\in\Delta^{m-1}}\;b'\left(M\pi-p\right).
\]
The objective on the left-hand side is affine in each of $b,\pi,$ and both feasible sets are compact and convex.  Thus, by the minimax theorem we can exchange the order of optimization.  However, the maximum of $b'(M\pi-p)$ over $b$ in the $\ell_{1}$ ball is precisely the dual norm $\left\Vert M\pi-p\right\Vert _{\infty}$, from which we see that the arbitrage profit is equal to
\[
\min_{\pi\in\Delta^{m-1}}\left\Vert M\pi-p\right\Vert _{\infty}=\ell_{\infty}(p),
\]
as claimed in Section \ref{sec:setup}.

\section{Prompts}\label{app:materials}

Queries share a common system message: ``You are a probabilistic forecaster. Estimate the probability of the requested event using the information supplied.''  The user message then presents, in order: the forecast timing (``You are making this forecast immediately after market close on \{date\}.''); the normalization (``each of the previous 60 daily simple returns has been divided by the sample standard deviation of those same returns,'' along with the standard deviation printed to eight decimals and the note ``The returns have not been mean-centered.''); the sixty dated variance-normalized returns, one per line; the bins (``B0 -- below $-1$ ($z<-1$)'' through ``B3 -- at least 1 ($z\geq1$)''); the event question; and the response instruction ``Return only a JSON object with one field named `probability'. Its value must be a number between 0 and 1, reported to four decimal places.'' Event questions take the form ``What probability would you assign to the following occurring? -- On future trading day 1, $z$ falls in B0 (below $-1$).'', with ``does not fall'' for complements and, for cross-day conjunctions, a ``both of the following'' lead with one line per day (``all of the following'' for three or more days).

\paragraph{Two-asset queries.} Two-asset questions show both assets' scales, define the scaled returns  $z_A$ and $z_B$, and display the sixty aligned dates as a three-column table (date, Asset A $z_A$, Asset B $z_B$); assets are labeled A and B only.  Marginal questions about each asset separately follow the single-asset form (``Asset A's $z_A$ falls in B0 (below $-1$)'') while joint-cell questions ask ``What probability would you assign to both of the following occurring on the same next trading day?'' with one bullet per asset.

\paragraph{Information arms.} The statistics arm appends ``Observed frequencies over the previous 60 trading days:'' with one count and percentage per bin, computed from the displayed history.  The instructions arm appends: ``Use the following forecasting procedure: 1. Count how many of the previous returns fall in each B range. 2. Use the resulting proportions as starting estimates for those ranges on each future trading day. 3. Use the ordered return history and your probabilistic judgment to estimate the requested event.''  The identity arm adds a line naming the company and ticker, while the headline arm further appends ``The following company headline was published before today's market close:'' with the verbatim pre-close headline.

\paragraph{Grouped arms.} Grouped arms present the same history, then ask ``Assign a probability to each of these labeled forecast events:'' with events listed as Q01, Q02, \ldots{} in an order randomized in each case, and request a JSON object with exactly those keys.  Complement and relation-poor pairs group items as described in the text, while the full-panel arm lists all fourteen events in one query.

\paragraph{Red herring arms.} Red herring arms insert between the history and the event question, ``Additional note from the person requesting this forecast:'' followed by one of seven fixed texts: Neutral personal detail (``I had coffee this morning, by the way.''), Investment stake (``I've got about 10\% of my savings invested in this company, by the way.''), Anxiety (``I'm honestly pretty scared about what this stock might do next.''), Gambling preference (``I love gambling and taking big risks, so this kind of forecast is exciting to me.''), Positive emotion (``I'm feeling really optimistic about this.''), Unrelated numeric anchor (``My phone battery is at 70\%, by the way.''), or Uninformed forecast anchor (``I don't know much about forecasting, but my gut says the probability is around 70\%'').

\paragraph{Unresolved outcome arms.} We show that coherence evaluations can also be applied to unresolved outcomes by eliciting forecasts for stock returns several years after the anchor dates. 

Specifically for $s \in \{2, 5, 10\}$ we replace the elicitation portion of the baseline prompt with ``The target is the stock's one-day simple return on the first NYSE trading session strictly after the \{s\}-year anniversary of \{date\}. This is not the stock's cumulative return over the intervening \{s\} years. Use the scale above for the target-session return. Treat every requested probability as conditional on the stock's survival through this horizon, meaning that it remains listed and has a valid one-day return on the target session. Do not assign probability mass to delisting or non-survival. On that target trading session, place \(z\) into one of the previously defined ranges. Conditional on the stock's survival through that horizon, what probability would you assign to the following occurring? On the target trading session, \(z\) falls in \{requested\_range\}.''

We find the mean incoherence statistics are \numFutureHorizonTwoYearFivePassProfitMean{} $[\numFutureHorizonTwoYearFivePassProfitCiLow{}, \numFutureHorizonTwoYearFivePassProfitCiHigh{}]$ for the 2-year horizon, \numFutureHorizonFiveYearFivePassProfitMean{} $[\numFutureHorizonFiveYearFivePassProfitCiLow{}, \numFutureHorizonFiveYearFivePassProfitCiHigh{}]$ for the 5-year horizon, and \numFutureHorizonTenYearFivePassProfitMean{} $[\numFutureHorizonTenYearFivePassProfitCiLow{}, \numFutureHorizonTenYearFivePassProfitCiHigh{}]$ for the 10-year horizon. This exercise shows that coherence evaluations are valid even for unresolved outcomes.

\section{Elicitation}\label{app:accounting}

Each model was accessed through one provider. Groq was used for GPT-OSS-120B; DeepInfra was used for GPT-OSS-20B, Qwen3-235B-A22B, DeepSeek V4 Flash, and Mistral Small 3.2; Amazon Bedrock was used for Claude Sonnet 5, Claude Fable 5, Claude Opus 5, Claude Opus 4.8, and Claude Opus 4.7; Google Vertex was used for Gemini 3.5 Flash and Llama-4 Maverick; and Azure was used for GPT-5.6 Luna, GPT-5.6 Sol, and GPT-5.6 Terra. Where supported, requests were sent with temperature 1.0, a 8,192 token output limit, the provider's default reasoning setting, and a task-specific seed (with a different seed per pass in multi-pass settings). The GPT-5.6 and Claude providers did not support including temperature, and the Claude provider did not support including seeds.

A response was accepted only if it was a valid JSON with exactly the requested fields and each generated probability in $[0,1]$. If errors, refusals, truncations, or malformed responses were received, the query was retried. {Across the completed workhorse GPT-OSS-120B runs, there were 4,744 retry attempts across 4,114 requests out of 345,500, attributed to rate limits (2,459), provider/response errors (2,268), interrupted dispatches (7), connection errors (5), timeouts (4), and missing routing metadata (1).}

\section{Linked Elicitation}\label{app:dependence}

We evaluate incoherence across linked and dispersed event sets in experiments that hold the size of the event set fixed. Intuitively, ``linked'' event sets contain all the joint events needed to decompose one marginal event, whereas ``dispersed'' event sets spread the joint events across different marginals. We do this for both two assets and two days. In both cases, the linked event sets impose one additional equality between joint probabilities and marginal probabilities.

Let \(i,j\in\{0,1,2,3\}\) index the four return bins. For the two-asset analysis, each panel contains the four marginals \(P(z_A\in i)\). A ``linked'' panel further contains all four joint cells \(P(z_A\in i,z_B\in j)\) for a fixed bin \(i\), which completes the decomposition of the marginal \(P(z_A\in i)\). A ``dispersed'' panel instead further contains \(P(z_A\in i,z_B\in(i+s)\bmod 4)\) for each \(i\) and a fixed shift \(s\). Repeating the construction with each asset as the base produces eight linked and eight dispersed event sets.

For the two-day analysis, each panel contains the four marginals \(P(z_1\in i)\). For every bin, the candidate joint events are in the two branches \(P(z_1\in i,z_2\in i)\) and \(P(z_1\in i,z_2\notin i)\). A ``dispersed'' panel selects one branch for each bin, giving \(2^4\) selections for each base day. A ``linked'' panel selects both branches for one completed bin, neither branch for a different omitted bin, and one branch for each remaining bin.

\begin{figure}[t]
  \centering
  \includegraphics[width=\linewidth]{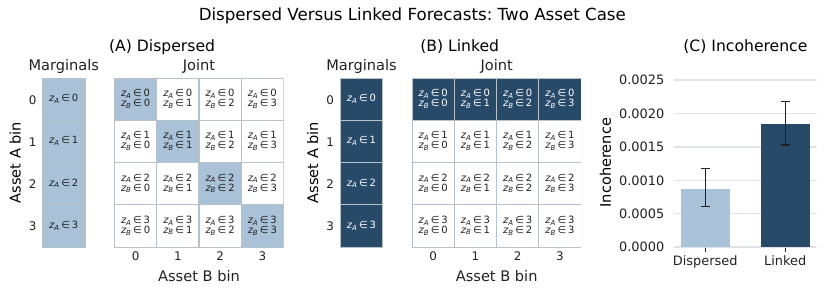}
  \caption{\textbf{Linking forecasts across two assets increases incoherence.} This figure reports the results of an experiment that holds the number of elicited events fixed, but varies the strength of relationships across events. Panels (A) and (B) include examples of dispersed and linked event sets (we include more permutations in the full analysis). Each event set contains four marginal events and four joint events. Panel (C) reports incoherence across the two elicitation strategies. Incoherence is on average $\numMatchedAssetsLinkedMinusDispersedProfitMeanRounded{}$ $[\numMatchedAssetsLinkedMinusDispersedProfitCiLowRounded{}, \numMatchedAssetsLinkedMinusDispersedProfitCiHighRounded{}]$ higher for linked event sets.}
  \label{fig:two_assets}
\end{figure}

\begin{figure}[t]
  \centering
  \includegraphics[width=\linewidth]{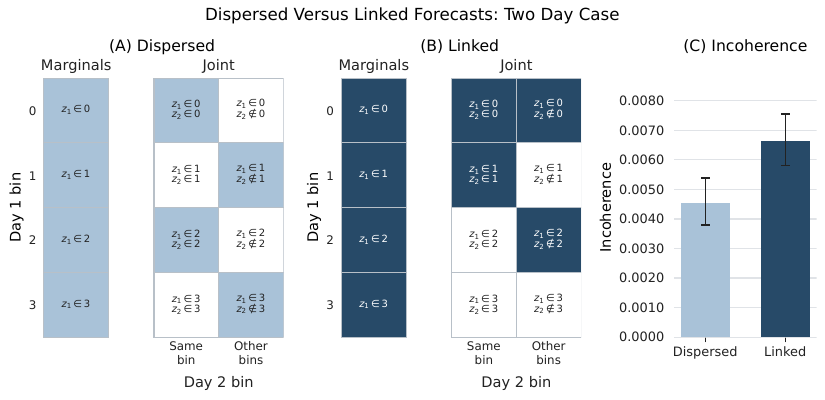}
  \caption{\textbf{Linking forecasts across two days increases incoherence.} This figure reports the results of an experiment that holds the number of elicited events fixed, but varies the strength of relationships across events. Panels (A) and (B) include examples of dispersed and linked event sets (we include more permutations in the full analysis). Each event set contains four marginal events and four joint events. Panel (C) reports incoherence across the two elicitation strategies. Incoherence is on average $\numMatchedHorizonsLinkedMinusDispersedProfitMeanRounded{}$ $[\numMatchedHorizonsLinkedMinusDispersedProfitCiLowRounded{}, \numMatchedHorizonsLinkedMinusDispersedProfitCiHighRounded{}]$ higher for linked event sets.}
  \label{fig:two_days}
\end{figure}

\section{Acknowledgements}
We thank Peter G. Chang, Jiafeng Chen, Ashesh Rambachan, and Mirac Suzgun for helpful comments and feedback. During the preparation of the manuscript, we used AI tools to generate code, and to draft and review portions of the manuscript. All AI outputs were reviewed by the authors.

\end{document}

%% file: paper_numbers.tex
\newcommand{\numBaselineHOneFivePassProfitMean}{0.002067}
\newcommand{\numBaselineHOneFivePassProfitCiLow}{0.001471}
\newcommand{\numBaselineHOneFivePassProfitCiHigh}{0.002737}
\newcommand{\numNBaselineHOneStockDaysProfitAboveTenBps}{48}
\newcommand{\numNBaselineHOneStockDaysProfitAboveFiftyBps}{15}
\newcommand{\numNBaselineHOneStockDaysProfitAboveOneHundredBps}{6}
\newcommand{\numFutureHorizonTwoYearFivePassProfitMean}{0.0080}
\newcommand{\numFutureHorizonTwoYearFivePassProfitCiLow}{0.0064}
\newcommand{\numFutureHorizonTwoYearFivePassProfitCiHigh}{0.0099}
\newcommand{\numFutureHorizonFiveYearFivePassProfitMean}{0.0070}
\newcommand{\numFutureHorizonFiveYearFivePassProfitCiLow}{0.0056}
\newcommand{\numFutureHorizonFiveYearFivePassProfitCiHigh}{0.0088}
\newcommand{\numFutureHorizonTenYearFivePassProfitMean}{0.0077}
\newcommand{\numFutureHorizonTenYearFivePassProfitCiLow}{0.0060}
\newcommand{\numFutureHorizonTenYearFivePassProfitCiHigh}{0.0096}
\newcommand{\numCrossModelMistralSmallThreePointTwoProfitMean}{0.1994}
\newcommand{\numCrossModelClaudeOpusFiveBrierMean}{0.1966}
\newcommand{\numCrossModelMistralSmallThreePointTwoBrierMean}{0.2157}
\newcommand{\numCrossModelProfitBrierSpearmanCorrelation}{0.9143}
\newcommand{\numNModelPairProfitCisExcludeZeroUnadjusted}{101}
\newcommand{\numNModelPairBrierCisExcludeZeroUnadjusted}{32}
\newcommand{\numTwoAssetFullMinusBothMarginalsProfitMean}{0.00244}
\newcommand{\numTwoAssetFullMinusBothMarginalsProfitCiLow}{0.001971}
\newcommand{\numTwoAssetFullMinusBothMarginalsProfitCiHigh}{0.002961}
\newcommand{\numMatchedAssetsLinkedMinusDispersedProfitMeanRounded}{0.00097}
\newcommand{\numMatchedAssetsLinkedMinusDispersedProfitCiLowRounded}{0.00081}
\newcommand{\numMatchedAssetsLinkedMinusDispersedProfitCiHighRounded}{0.0012}
\newcommand{\numMatchedHorizonsLinkedMinusDispersedProfitMeanRounded}{0.0021}
\newcommand{\numMatchedHorizonsLinkedMinusDispersedProfitCiLowRounded}{0.0019}
\newcommand{\numMatchedHorizonsLinkedMinusDispersedProfitCiHighRounded}{0.0023}